\documentstyle[12pt]{article}

\def\appendix#1{
  \addtocounter{section}{1}
 \setcounter{equation}{0}
  \renewcommand{\thesection}{\Alph{section}}
 \section*{Appendix \thesection\protect\indent \parbox[t]{11.715cm} {#1}}
  \addcontentsline{toc}{section}{Appendix \thesection\ \ \ #1}
  }

\def\bea{\begin{eqnarray}}
\def\eea{\end{eqnarray}}
\def\be{\begin{equation}}
\def\ee{\end{equation}}

\begin{document}
\thispagestyle{empty}

\begin{center}
{\huge \bf On Pure Lattice Chern-Simons Gauge Theories}\\
\vskip 0.5truein 
F. Berruto$^{(a)}$, M.C. Diamantini$^{(b) {}^*}$ and P. Sodano$^{(a)}$
\vskip 0.3truein
a){\it Dipartimento di Fisica and Sezione
I.N.F.N., Universit\`a di Perugia, Via A. Pascoli I-06123 Perugia,
Italy}\\
\vskip 0.3truein
b){\it Department of Theoretical Physics, Oxford University, 1 Keble Rd,
Oxford UK} 
\vskip 0.5truein
DFUPG-19-00
\vskip 1.0truein
{\bf Abstract}
\end{center}
\vskip 0.3truein
\addtocounter{footnote}{1}
\footnotetext{Supported by a Swiss National Science Foundation fellowship.}

We revisit the lattice formulation of the Abelian Chern-Simons model defined
on an 
infinite Euclidean lattice. We point out that any gauge invariant, local and
parity odd 
Abelian quadratic form exhibits,
in addition to the zero 
eigenvalue associated with the gauge invariance and to the physical zero mode
at $\vec{p}=\vec{0}$ 
due to translational invariance, a set of extra zero eigenvalues inside the
Brillouin zone. 
For the Abelian Chern-Simons theory, which is linear in the derivative, this
proliferation of zero modes 
is reminiscent of the Nielsen-Ninomiya no-go theorem for fermions. 
A gauge invariant, local and parity even term such as the Maxwell action
leads to the 
elimination of the extra zeros by opening a gap with a mechanism similar to
that leading to Wilson fermions on the lattice.

\newpage
\setcounter{page}1
\setcounter{equation}0

It is by now well known that in odd space-time dimensions, there is the
possibility of adding a gauge 
invariant, topological Chern-Simons (CS) term to the gauge field action. 
The CS term breaks 
both the parity and time-reversal symmetries and, when coupled with a Maxwell or
Yang-Mills term, leads to massive gauge excitations~\cite{jackiw}. 
For an Abelian model in three space-time dimensions, the pure CS 
Lagrangian is defined as
\begin{equation}
{\cal L}_{CS}=\frac{k}{2}A_{\mu}\epsilon^{\mu \alpha
\nu}\partial_{\alpha}A_{\nu}\quad ,
\label{lagrangian1}
\end{equation} 
where $k$ is a dimensionless coupling constant. 

The pure CS theory is a 
topological field theory~\cite{witt}. It is exactly solvable and it is used to
compute topological invariants 
of three manifolds, the knot invariants for links
embedded in 
three-manifolds~\cite{witt}.
As a model for physical phenomena, being dominant at large distances, the CS
action may be used as a low energy
effective field theory for condensed matter systems such as the fractional
quantum Hall effect~\cite{hall}
or Josephson junction arrays~\cite{cris}. 
Of great interest  is also the relationship of CS theories to conformal
field theories in 
two dimensions~\cite{conf}. 

While in the continuum the pure CS theory is exactly solvable, things are quite
different on the lattice.
As originally shown by Fr\"{o}hlich and Marchetti~\cite{marchetti},
the kernel defining the CS action  exhibits a set of zeroes which are not 
due to gauge invariance; thus, the theory is not integrable even after gauge
fixing. 
The action (\ref{lagrangian1}) is of first order in the derivatives, and the
appearence of extra zeros in its lattice formulation is reminescent of the
``doubling'' of fermions on the lattice~\cite{doubling, wilson}. While, for
fermion models, the doubling made
for many years impossible to
formulate chiral gauge theories on the lattice (for recent development on this
subject see~\cite{martin0}), the extra zeroes of the CS
action make only the definition of a parity odd theory on the lattice sick.
In the following we revisit the lattice formulation of the
Euclidean version of the 
Abelian CS model defined by the Lagrangian (\ref{lagrangian1}). 
Previous studies of pure CS theory on the lattice  have been carried out
using
 the Hamiltonian formalism in~\cite{seme},
by introducing a mixed CS action with two gauge fields with opposite parity
or by means of two gauge fields living
on the links of two dual lattices (thereby obtaining in both cases a parity
even action)~\cite{kantor, adams}.

In this letter
we shall show that, due to the Poincar\'e lemma recently proved
on the lattice by L\"{u}scher~\cite{luscher}, the non-integrability of 
the CS kernel is a general feature of 
any gauge-invariant, local and parity odd gauge theory on a lattice.
Moreover, again as a consequence of the Poincar\'e lemma, we shall show
that the only
gauge invariant regularization of the CS action may be obtained by adding to
it a parity even term; of course, the most physical choice is the Maxwell term. 
Since the addition of a Maxwell term regularizes the CS action, the presence
of the extra 
zeros in the CS action did not cause any problem
in previous investigations of the Maxwell-CS action on the
lattice~\cite{marchetti,
luscher2, pasquale}. 

We consider an infinite Euclidean cubic lattice with lattice spacing $a$,
which we set to unity ($a=1$). We shall denote a lattice 
site by the vector $\vec{x}$ and a link between $\vec{x}$ and
$\vec{x}+\hat{\mu}$ ($\mu=0,1,2$) by $(\vec{x},\hat{\mu})$. 
Forward and backward difference operators are given by
$d_{\mu}f(\vec{x})=f(\vec{x}+\hat{\mu})-f(\vec{x})=
(S_{\mu}-1)f(\vec{x})$ and $\hat{d}_{\mu}f(\vec{x})
=f(\vec{x})-f(\vec{x}-\hat{\mu})=(1-S_{\mu}^{-1})f(\vec{x})$, where
$S_{\mu}f(\vec{x})=f(\vec{x}+\hat{\mu})$, $S^{-1}_{\mu}f(\vec{x})=
f(\vec{x}-\hat{\mu})$ are the forward and backward shift operators respectively.
 Summation by parts on the lattice interchanges the forward and backward
 derivatives:
$\sum_{\vec{x}}f(\vec{x})d_{\mu}g(\vec{x})
=-\sum_{\vec{x}}\hat{d}_{\mu}f(\vec{x}) g(\vec{x})$.

The lattice Fourier transformation of the 
gauge field $A_{\mu}$ is given by
\begin{equation}
A_{\mu}(\vec{x})=\int_{\cal{B}} \frac{d^3 p}{(2\pi)^3} e^{-i
\vec{p}\cdot\vec{x}} e^{-ip_{\mu}/2}A_{\mu}(\vec{p})\quad .
\label{fourier1}
\end{equation}
Due to the phase factor $e^{-ip_{\mu}/2}$, $A_{\mu}(\vec{p})$ is
antiperiodic if 
$p_{\mu}\longrightarrow p_{\mu}+2\pi (2n+1)$, with $n$ integer. 
The integration over momenta in eq.(\ref{fourier1}) 
is restricted to the Brillouin zone 
${\cal B}=\left\{p_{\mu}| -\pi\le p_{\mu}\le \pi \ ,\ \mu =0,1,2\right\}$. 
Under parity, which on an Euclidean cubic lattice corresponds to the
simultaneous
 inversion of all three directions,  
$A_{\mu}(\vec{x})\longrightarrow A_{\mu}(-\vec{x}-\hat{\mu})$ and
 $A_{\mu}(\vec{p})\longrightarrow -A_{\mu}(-\vec{p})$.

The CS action on an Euclidean lattice derived by Fr\"{o}lich 
and Marchetti~\cite{marchetti} is:
\begin{equation}
S=\sum_{\vec{x}}A_{\mu}(\vec{x})\tilde{K}_{\mu \nu}(\vec{x} -
\vec{y})A_{\nu}(\vec{y})\quad ,
\label{action1}
\end{equation}
where $\tilde{K}_{\mu\nu}=K_{\mu\nu}+\hat{K}_{\mu\nu}$, and
\begin{eqnarray}
K_{\mu\nu}(\vec{x} -
\vec{y})&=&S_{\mu}^{\vec{y}}\ \epsilon_{\mu\alpha\nu}\ d_{\alpha}^{\vec{y}}\
\delta_{\vec{x},\vec{y}}\label{ker1}\quad ,\\
\hat{K}_{\mu\nu}(\vec{x} - \vec{y})&=&S_{\nu}^{-1,\vec{y}}\ \epsilon_{\mu
\alpha\nu}\ \hat{d}_{\alpha}^{\vec{y}}\ \delta_{\vec{x},\vec{y}}\label{ker2}\quad .
\end{eqnarray}
$K$ and $\hat{K}$ are exchanged by summation by parts. 

Both $K$ and $\hat{K}$  define  a gauge invariant and parity odd 
kernel. In momentum space, apart from the zero due to gauge invariance, $K$
 and $\hat{K}$ have eigenvalues 
$\lambda(p)=\hat{\lambda}^{\dagger}(p)=\pm 2
e^{-i\sum_{\mu=0}^2p_{\mu}/2}\sqrt{\sum_{\mu=0}^2\sin^2 p_{\mu}/2}$
and exhibit no extra zeroes apart from the one at zero momentum, associated
with translational invariance. 
The kernels (\ref{ker1}) and (\ref{ker2}), after gauge fixing,
both define an integrable CS action. However,
since (\ref{action1}) is a quadratic form in $A_\mu(\vec{x})$, the kernel 
should be symmetric  under the simultaneous exchange of $\mu
\rightarrow \nu, (\vec{x}) \rightarrow (\vec{y})$ (according to~\cite{karsten},
we call this Bose symmetry); it is easy to see that only the linear combination 
$\tilde{K}=K+\hat{K}$ respects this symmetry and thus provides an acceptable
definition of the lattice CS action. This has far reaching
consequences on the integrability of the action (\ref{action1}) since, in
momentum
space, the operator $\tilde{K}(p)=K(p)+\hat{K}(p)$ has, apart from the zero
mode associated with gauge invariance, 
eigenvalues given by
\begin{equation}
\tilde{\lambda}(p)=\pm 2\sqrt{1+\cos\sum_{\mu
=0}^2p_{\mu}}\sqrt{3-\sum_{\mu=0}^2\cos p_{\mu}}\quad .
\label{spectrum1}
\end{equation}
One gets $\tilde{\lambda}=0$ whenever $\cos\sum_{\mu=0}^2p_{\mu}=-1$, $i.e.$
when $\sum_{\mu=0}^2p_{\mu}=
(2n+1)\pi$,  which defines planes of zeros with co-dimension 1 in the Brillouin
zone: the CS action (\ref{action1}) is thus not integrable. 
The properties of $K$ and $\hat{K}$ parallel the ones  of the forward and
backward derivatives, which in 
momentum space read $d_{\mu}\rightarrow e^{ip_{\mu}/2}\hat{p}_{\mu}$ and 
$\hat{d}_{\mu}\rightarrow e^{-ip_{\mu}/2}\hat{p}_{\mu}$ 
with $\hat{p}_{\mu}=2\sin p_{\mu}/2$: they do not have extra zeroes inside 
the Brillouin zone, but their linear combination 
$d + \hat{d} \rightarrow 2 \cos(p_\mu/2) \hat{p}_\mu $ has zeros at the
border of the Brillouin zone $p_\mu = \pm
\pi$.

The appearance of the extra zeroes is not due to the specific form of the kernel
in (\ref{action1}). In fact,
let us consider an action given by
\begin{equation}
S=\sum_{\vec{x},\vec{y}}A_{\mu}(\vec{x})G_{\mu
\nu}(\vec{x}-\vec{y})A_{\nu}(\vec{y})\quad .
\label{action2}
\end{equation}
We shall now prove that, under the assumptions that

\noindent i) (\ref{action2}) is local on the lattice;

\noindent ii) (\ref{action2}) is gauge invariant: $\hat{d^x_{\mu}}G_{\mu
\nu}(\vec{x}-\vec{y})
=\hat{d^y_{\nu}}G_{\mu \nu}(\vec{x}-\vec{y})=0$;

\noindent iii) (\ref{action2}) is odd under parity;

\noindent (\ref{action2}) is not integrable.

The fact that $S$ is a quadratic form implies that 
$G_{\mu \nu}(\vec{x}-\vec{y})=G_{\nu \mu}(\vec{y}-\vec{x})$ (Bose symmetry).
By Fourier transforming eq.(\ref{action2}), one gets
\begin{equation}
\int_{\cal B} \frac{d^3p}{(2\pi)^3}A_{\mu}(-\vec{p})\tilde{G}_{\mu
\nu}(\vec{p})A_{\nu}(\vec{p})
\label{action3}
\end{equation}
with $\tilde{G}_{\mu \nu}(\vec{p})=e^{ip_{\mu}/2}G_{\mu
\nu}(\vec{p})e^{-ip_{\nu}/2}$
(no sum over $\mu$ and $\nu$).
Note that, in order for ${G}_{\mu \nu}(\vec{p})$ to be a periodic function
of the
momentum $\vec{p}$, $\tilde{G}_{\mu \nu}(\vec{p})$ must be antiperiodic in
$p_\mu \rightarrow p_\mu + 2\pi(2n+1)$ and 
$p_\nu \rightarrow p_\nu + 2\pi(2n+1)$, with $n$ integer. Moreover, 
$\tilde{G}_{\mu \nu}(\vec{p})$ must be a periodic function of the component
of the three-momentum different from $\mu$ and $\nu$, since no extra phase is 
present in the definition of  $\tilde{G}_{\mu \nu}(\vec{p})$.
These properties of periodicity and antiperiodicity are crucial in the proof
of the non-integrability of (\ref{action2}).

The kernel 
$\tilde{G}_{\mu \nu}(\vec{p})$ is such that 
\begin{equation}
\tilde{G}_{\mu \nu}(\vec{p})=\tilde{G}_{\nu \mu}(-\vec{p})
\label{kbose}
\end{equation}
due to the Bose symmetry. In momentum space the assumptions i)-iii) imply 
that $\tilde{G}_{\mu \nu}(\vec{p})$ is an analytic function (from
locality) satisfying to
\begin{equation}
\hat{p}_{\mu}\tilde{G}_{\mu \nu}(\vec{p})= 0
\label{kge}
\end{equation}
from gauge invariance, and to
\begin{equation}
\tilde{G}_{\mu \nu}(\vec{p})=-\tilde{G}_{\nu \mu}(\vec{p}) 
\label{kodd}
\end{equation}
from parity oddness.
As a consequence of (\ref{kge}), using the Poincar\'e lemma, one may write
\begin{equation}
\tilde{G}_{\mu \nu}(\vec{p})= 
\epsilon_{\rho \mu \nu} \hat{p}_{\rho}f(\vec{p}) ,
\label{kgi}
\end{equation}
where $f(\vec{p})$ is an analytic function of $\vec{p}$.
Since $\tilde{G}_{\mu \nu}(\vec{p})$ must be periodic in the component of
the momentum different from $\mu$ and $\nu$, one has that $f(\vec{p})$
must be antiperiodic in $p_\rho \rightarrow p_\rho + 2\pi(2n+1)$.
Moreover, since all the dipendence of $\tilde{G}_{\mu \nu}(\vec{p})$ on 
$p_\mu$ and $p_\nu$ is carried by $f(\vec{p})$, the latter must also be
antiperiodic 
with respect to $p_\mu$ and $p_\nu$:
in order for the kernel $G_{\mu \nu}(\vec{p})$ to be a periodic function of
the momenta, $f(\vec{p})$ 
must satisfy to
\begin{equation}
f(p_{0}+2\pi (2n+1),p_{1}+2\pi (2n+1),p_{2}+2\pi
(2n+1))=-f(p_0,p_1,p_2)\quad .   
\label{fodd}     
\end{equation}
Due to eqs.(\ref{kodd},\ref{kgi}) one has that 
$f(\vec{p})=f(-\vec{p})$. 
 From eq.(\ref{fodd}), one may easily check that for $p_0=p_1=p_2=\pm \pi$ 
one gets
\begin{equation}
f(\pm \pi,\pm \pi,\pm \pi)=-f(\pm \pi,\pm \pi,\pm \pi)=0\quad 
\label{ff0}
\end{equation}
(this is also true when any two component of the momentum are equal to zero and 
one is equal to $\pm \pi$).
Since the spectrum of $G_{\mu \nu}(\vec{p})$ is given by
$G(\vec{p})=\pm |f(\vec{p})|\sqrt{\sum_{\mu=0}^2 \hat{p}^2_{\mu}}$
(apart from the zero due to gauge invariance),
 eq.(\ref{ff0}) implies that the kernel 
$G(\vec{p})$ 
exhibits extra zeroes at the edges of the Brillouin zone 
and is thus not integrable.
Note that the presence of the extra zeros is completely independent on the
nature,
complex or real, of the function $f(\vec{p})$. This observation is pertinent 
since, in Euclidean space-time, the pure CS action is purely immaginary.

Relaxing the assumption iii) one may study the general form of a
gauge invariant local action in three dimensions.
With the help  of the Poincar\'e lemma~\cite{luscher} it is easy to show that
the kernel $\tilde{G}_{\mu \nu}(\vec{p})$ can be divided into the sum of parity
even and parity odd terms. Since, due to locality, $\tilde{G}_{\mu
\nu}(\vec{p})$ is an
analytic function of $\vec p$, it may be expanded in Taylor series: all the
terms having even power of the momenta are parity even, while the terms with
odd power of the momenta are parity odd.
The terms with the lowest number of derivatives in this expansion are the
CS term defined in~\cite{marchetti} and the Maxwell term, whose kernel on
the lattice
is:
\begin{equation}
M_{\mu \nu}=-\Box\delta_{\mu \nu}+d_{\mu}\hat{d}_{\nu}=K_{\mu
\rho}\hat{K}_{\rho \nu} ,
\label{max}
\end{equation}
where $\Box=\sum_{\mu=0}^2d_{\mu}\hat{d}_{\mu}$ 
is the Laplacean in three dimensions.
Since all the parity odd terms fullfill the assumptions of the theorem
they generate extra zeroes in the spectrum. The only gauge invariant way to
 regularize the CS action is then the addition of a 
parity even term such as the Maxwell term. 

For the Maxwell-CS
theory on the 
lattice the kernel $\Gamma_{\mu \nu}$ may be written as 
\begin{equation}
\Gamma_{\mu \nu}=\frac{1}{4 e^2}M_{\mu \nu}
+i k G_{\mu \nu}.
\label{kmcs}
\end{equation}
In (\ref{kmcs}) $k$ is
dimensionless and $e^2$ has the dimension of a mass; the Maxwell term is an
irrelevant operator and the CS action dominates in the infrared region.
The Fourier transform of $\Gamma_{\mu \nu}$, apart from a zero mode due to
gauge 
invariance, has eigenvalues given by
\begin{equation}
\lambda_{MCS}(\vec p)=\frac{1}{2e^2}\sum_{\mu =0}^2(1-\cos p_{\mu})+
i k G(\vec p)\quad ,
\label{eigenmcs}
\end{equation}
and, as it stands, it is free from extra zeroes in the Brillouin zone since
the first term in (\ref{eigenmcs}), which is the Fourier transform 
of the Maxwell kernel, is zero only at zero momentum, and at the corners of the
Brillouin zone, $p_{\mu}=\pm \pi,\ \mu=0,1,2$, takes the value 
$\lambda_{MCS}(\vec p) =3/e^2$. 

Since the CS action is purely immaginary, the addition of the Maxwell term 
is used also in the continuum theory to provide a proper definition of the
functional
integral in the partition function of the pure CS theory. The CS limit is
reached also there
by taking the limit $e^2 \longrightarrow \infty$ after Gaussian integration.

The regularization of the extra zeros in the CS action by adding a Maxwell term
and thereby opening a gap in the fermion spectrum is similar to the 
mechanism of the Wilson fermion where a gap is opened and the energy does not
have secondary minima at the non-zero corners of the Brillouin zone.
As in the case of the Wilson fermions~\cite{kogwil}, the regularization is
done by means of an
irrelevant operator and the continuum limit $a \longrightarrow 0$ is
not changed by this addition.
Moreover, as the Wilson action explicitly breaks chiral symmetry, the action
obtained after the addition of the Maxwell term is not anymore defined under
parity.

A key result of our analysis is a  no-go theorem in the lattice
regularization of the 
pure CS theory, if one requires locality, gauge invariance and parity oddness
on the lattice.
Clearly all the arguments needed for the proof rely on the definition of the
parity transformation
for the gauge field, i.e. on $A_{\mu}(\vec{x})\longrightarrow
A_{\mu}(-\vec{x}-\hat{\mu})$.
One possible way out is to define a new parity transformation under which
the CS action is still odd but the kernel is integrable.
The new definition of the parity transformation should then play a role
analogous to the
Ginsparg-Wilson relation~\cite{wilson1} in the definition of 
chiral gauge theories on the lattice. When 
the Ginsparg-Wilson relation holds, the lattice fermion action 
has an exact chiral symmetry~\cite{martin} and the no-go theorem of 
Nielsen and Ninomiya~\cite{nielsen} is avoided.
It is pertinent to point out that, while for chiral gauge theories the
problem of the non-perturbative 
regularization of the theory in a way preserving the symmetry is related 
to the presence, at the quantum level, of anomalies, for the CS action
the theory in the continuum is well defined and solvable.
This hints to the fact that the non-integrability on the lattice of the CS
action is 
only a lattice artifact, due mainly to the existence of  
two derivatives giving rise to the two kernels $K$ and $\hat K$.

\vskip 0.3 truein
\noindent
{\bf Acknoledgments} 
\noindent
We are particularly in debt with M. L\"uscher for many helpful discussions.
 We  also  greatly benefited from conversation on topics relevant to 
the subject of this paper with G. Grignani, G.W. Semenoff and
C. A. Trugenberger. M.C.D. and F.B. thank the 
Theory Division of C.E.R.N., where part of this work was performed, for 
the warm hospitality. This work was partially supported by grants from 
M.U.R.S.T. and I.N.F.N..


\end{document}